\documentclass[useAMS,usenatbib]{mn2e}

\usepackage{epsfig}
\usepackage{amsmath}
\usepackage{amssymb}
\usepackage{natbib}

\usepackage[dvips]{color}

\newcommand{\be}{\begin{eqnarray}} 
\newcommand{\ee}{\end{eqnarray}} 
\newcommand{\simprop}{\buildrel \sim \over \propto}
\usepackage{epstopdf}

\usepackage{multicol}

\newcommand{\lum}{\mathrm{erg~s}^{-1}}

\newcommand{\mdot}{\mathrm{M_{\odot}~yr}^{-1}}

\newcommand{\averagedmdot}{$\langle \dot{M} \rangle$}

\def \mnras {MNRAS}
\def \apj {ApJ}
\def \apjs {ApJS}
\def \apjl {ApJL}
\def \aap {A\&A}
\def \nat {Nature}

\def \pasj {PASJ}

\def \prc {PhRvC}

\title[Quiescent neutron-star VFXTs]{Testing the deep-crustal heating model using quiescent neutron-star very-faint X-ray transients and the possibility of partially accreted crusts in accreting neutron stars}
\author[Wijnands, Degenaar, \& Page]
{R. Wijnands$^{1}$\thanks{e-mail: r.a.d.wijnands@uva.nl},  
N. Degenaar$^{2}$\thanks{Hubble Fellow},
D. Page$^{3}$
\\
$^{1}$Astronomical Institute ``Anton Pannekoek'', 
University of Amsterdam, 
Postbus 94249, 1090 GE Amsterdam, the Netherlands\\
$^{2}$Department of Astronomy, University of Michigan, 500 Church Street, Ann Arbor, MI 48109-1042, USA\\
$^{3}$Instituto de Astronom\'\i a, Universidad Nacional Aut\'onoma de M\'exico, Mexico D.F. 04510,Mexico\\
}

\begin{document}


\pagerange{\pageref{firstpage}--\pageref{lastpage}} \pubyear{0000}

\maketitle

\label{firstpage}

\begin{abstract} 
It is assumed that accreting neutron stars in low-mass X-ray binaries
are heated due to the compression of the existing crust by the freshly
accreted matter which gives rise to a variety of nuclear reactions in
the crust.  It has been shown that most of the energy is released deep
in the crust by pycnonuclear reactions involving low-Z elements (the
deep-crustal heating scenario). In this paper we discuss if neutron
stars in so-called very-faint X-ray transients (VFXTs; those which
have outburst peak 2-10 keV X-ray luminosities $< 1\times 10^{36}$ erg
s$^{-1}$) can be used to test this deep-crustal heating
model. We demonstrate that such systems would indeed be very interesting objects to test the deep-crustal heating model with, but that the interpretation of the results might be challenging 
because of the large uncertainties in our estimates of the accretion
rate history of those VFXTs, both the short-term (less than a few tens
of thousands of years) and the one throughout their lifetime. The
latter is particularly important because it can be so low that the
neutron stars might not have accreted enough matter to become massive
enough that enhanced core cooling processes become active. Therefore,
they could be relatively warm compared to other systems for which such
enhanced cooling processed have been inferred. However, the amount of
matter can also not be too low because then the crust might not have
been replaced significantly by accreted matter and thus a hybrid crust
of partly accreted and partly original, albeit further compressed
matter, might be present. This would inhibit the full range of
pycnonuclear reactions to occur and therefore possibly decreasing
the amount of heat deposited in the crust. More detailed calculations
of the heating and cooling properties of such hybrid crusts have to be
performed to be conclusive. Furthermore, better understanding is
needed how a hybrid crust affects other properties such as the thermal
conductivity. A potential interesting way to observe the effects of a hybrid crust on the heating and cooling of an accreting neutron star is to observe the crust cooling of such a neutron star after a prolonged (years to decades) accretion episode and compare the results with similar studies performed for neutron stars with a fully accreted crust.  We also show that some individual neutron-star low-mass
X-ray binaries might have hybrid crusts as well as possibly many of
the neutron stars in high-mass X-ray binaries. This has to be taken
into account when studying the cooling properties of those systems
when they are in quiescence. In addition, we show that the VFXTs are
likely not the dominate transients that are associated with the
brightest ($\sim 10^{33}$ erg s$^{-1}$) low-luminosity X-ray sources in
globular clusters as was previously hypothesized.
\end{abstract}

\begin{keywords}
X-rays: binaries - binaries: close 
\end{keywords}

\section{Introduction}

Neutron-star low mass X-ray binaries (LMXBs) harbor neutron stars
which are accreting matter from a close-by low-mass (typically $<1$
M$_\odot$) companion star which transfer mass due to Roche-lobe
overflow.  In most systems the neutron star does not continuously
accrete matter. Generally, those systems (called X-ray transients) are
in their quiescent state in which they do not accrete at all or only
at a very low rate. Only occasionally they exhibit bright X-ray
outbursts during which their X-ray luminosities increase by several
orders of magnitude. Such outbursts are most probably caused by a very
large increase in the mass accretion rate onto the neutron stars due
to instabilities in the accretion disc \citep[see the review
by][]{2001NewAR..45..449L}. The X-ray transients can be divided in
sub-groups based on their peak 2--10 keV X-ray luminosities in
outburst. In particular, a special sub-group has been called the
very-faint X-ray transients (VFXTs) that have peak luminosities
between $1\times 10^{34}$ and $1\times 10^{36}$ erg s$^{-1}$
\citep[see][]{2006A&A...449.1117W}, whereas the brighter transients
have peak luminosities of $10^{36-39}$ erg s$^{-1}$.

In quiescence, neutron-star transients can still be detected using
sensitive X-ray satellites, and it has been found that in many systems
a soft, most likely thermal, component is present with a typical
black-body temperature of 0.1--0.3 keV \citep[see, e.g.,][and
references to those
papers]{1987A&A...182...47V,1996PASJ...48..257A,1998A&ARv...8..279C,1999ApJ...514..945R}. In
addition, for many systems an additional spectral component above 2
keV has also been detected \citep[the non-thermal power-law component;
see, e.g.,][]{1996PASJ...48..257A,1999ApJ...514..945R}, which can even
dominate the 0.5--10 keV X-ray flux in some systems
\citep[e.g.,][]{2002ApJ...575L..15C,2004MNRAS.354..666J,2005ApJ...618..883W,2012arXiv1204.6059D}. The
origin of the non-thermal component is not clear \citep[see, e.g., the
discussions in][]{1998A&ARv...8..279C,2012arXiv1204.6059D} but it is
generally assumed that the soft component is the thermal emission from
the neutron star surface, either due to very-low level residual
accretion onto the surface or due to the cooling of the neutron star that
has been heated by the matter accreted during the outbursts.

During the accretion phases, matter accumulates on the surface of the
neutron star. This matter compresses the underlying layers of the
neutron star crust.  If the accretion continues long enough the
original catalyzed crust can be completely replaced by a new crust
made of accreted matter (\citealt{1979PThPh..62..957S},
\citealt{1990A&A...229..117H}).  The original crust is pushed down
into the neutron star until it fuses together with the core. The
composition of the accreted crust should be quite different from the
original, catalyzed crust, i.e., richer in low-Z elements
\citep{1990A&A...229..117H}. It has been postulated that when the
accreted matter sinks into the crust due to the compression induced by
freshly accreted material onto the star, a chain of non-equilibrium
reactions occurs in the crust that generates heat \citep[electron
captures, neutron drips and pycnonuclear
reactions;][]{1990A&A...227..431H,2003A&A...404L..33H,2008A&A...480..459H,2007ApJ...662.1188G}. Most
of the heat is released deep in the crust (at densities $>10^{12}$ g
cm$^{-1}$) due to pycnonuclear reactions involving low-Z
elements. This heat is conducted inwards, heating the core, and
outwards, where it is emitted as thermal emission from the surface.
This model has been called the ``deep crustal heating model''
\citep{1998ApJ...504L..95B}. This model has been tested by comparing the observed thermal emission of quiescent neutron stars with predictions based on estimated of their time-averaged accetion rates (see section~\ref{Sec:crustheating_data}). Another exciting possibility is to study the thermal relaxation of accretion-heated neutron star crusts after the end of accretion outbursts (see also section \ref{forward}).

In Section~\ref{Sec:crustheating} we briefly describe the deep crustal
heating model and compare the model with the available data. We also
calculate the time-scale on which the core reacts to changes in the
long-term averaged accretion rate. In Section~\ref{Sec:qLx} we
calculate, in the frame work of the deep-crustal heating model, the
expected quiescent luminosity of neutron-star VFXTs, in order to use
those systems to test the deep-crustal heating model. We argue that it
might be possible that during their life those systems might not have
accreted enough matter to have fully replaced their original
neutron-star crust with an accreted one which could significantly
inhibit the pycnonuclear heating reactions.  In
Section~\ref{sec:discussion} we discuss how the VFXTs can still be
used to test the model and also discuss potential other sources which
might harbor neutron stars with only partly accreted crusts.

\section{Quiescent neutron stars and the deep crustal heating model}
\label{Sec:crustheating}

In this heating/cooling model one can calculate the thermal state of
the neutron star with a simple energy balance consideration by writing
\be
\frac{dE_{th}}{dt} = C_V \frac{dT}{dt}  = H - L_\gamma - L_\nu
\label{Eq:cool}
\ee
where $E_{th}$ is the thermal energy of the neutron star, $C_V$ its
total specific heat, and $T$ its core temperature.  $H$ is the total
heating rate, and the two energy sinks are the star's thermal photon
luminosity, $L_\gamma$, and its neutrino luminosity $L_\nu$.

In the deep crustal heating scenario $H$ is taken as a time average
\be
\begin{array}{cc} 
\displaystyle
H \rightarrow \langle H \rangle = \langle \dot{M} \rangle {Q_{nuc}\over m_u} \approx
\\
\displaystyle
10^{33} { \langle \dot{M} \rangle \over 10^{-11}~ M_\odot ~{\rm yr}^{-1}}
{ Q_{nuc} \over 1.5 ~{\rm MeV}} {\rm ~erg~ s}^{-1}
\end{array}
\label{Eq:Heat} 
\ee
where $\langle \dot{M} \rangle$ is the long term time-averaged mass
accretion rate onto the neutron star, $Q_{nuc}$ the amount of heat,
per accreted nucleon, deposited in the crust and $m_u$ the atomic
mass unit.  Theoretical predictions (e.g.,
\citealt{2008A&A...480..459H}) obtain values for $Q_{nuc}$ between 1
and 2 MeV.

When $\langle \dot{M} \rangle$ has been stable for a long enough time
the neutron star is in thermal equilibrium (see
Section~\ref{Sec:tauth} for estimates of the thermal response
time-scale).  Hence, from Eq.~(\ref{Eq:cool}) with $dT/dt = 0$, one
obtains the expected $L_\gamma$, or the observable quiescent thermal luminosity $L_q$\footnote{
We will designate the photon luminosity by $L_\gamma$ or $L_q$, using the former when considering it
as an energy sink, as in Eq.~(\ref{Eq:cool}), and the latter when it is seen as an observed thermal luminosity.}
, as
\be
L_q = \langle H \rangle - \langle L_\nu \rangle \, .
\label{Eq:quiescent}
\ee
\cite{1998ApJ...504L..95B} showed that when $\langle L_\nu \rangle$ is negligible the very simple relation
\be
L_q = \langle H \rangle = \langle \dot{M} \rangle  Q_{nuc}/m_u
\label{Eq:LqMdot}
\ee
agreed with several observations of quiescent neutron-star LMXB
transients.  However, \cite{2001ApJ...548L.175C} emphasized that many
such systems are hot enough that $\langle L_\nu \rangle$ is not
necessarily negligible and could even be in a regime where $L_\nu \gg
L_\gamma$.  In the case that fast neutrino emission is possible (as one
would expect for a massive neutron star) this would explain the low
values of $L_q$ observed from several systems that are discrepant with
the simplified Eq.~(\ref{Eq:LqMdot}) since with a very large $L_\nu$
one could have $L_q \ll \langle \dot{M} \rangle Q_{nuc}/m_u$.

Neutrino emission processes can be roughly divided in two categories,
either ``slow'' or ``fast'' that differ in their temperature dependence
and efficiency
\citep{2004ARA&A..42..169Y,2006NuPhA.777..497P,2006ARNPS..56..327P}.
Slow processes include the modified Urca (``MUrca'') and several
bremsstrahlung processes whose emissivities can be roughly written as
$\epsilon_\nu^\mathrm{slow} = Q T_9^8$ erg cm$^{-3}$ s$^{-1}$ where
$T_9 \equiv T/10^9$ K.  Values for $Q$ range from $\simeq 10^{19}$ for
the bremsstrahlung processes to $10^{21}$ for the MUrca.  A higher
emissivity for the MUrca has been proposed by
\cite{1986ZhETF..90.1505V}, called the Medium-Modified Urca process
(``MMUrca''), resulting from the softening of the pion mode and that is a
precursor to the pion condensate.  The MMUrca has a $Q$ that reach
$10^{23}$ and it smoothly merges into the pion condensate emissivity
(see below) with growing density.  The fast processes have
emissivities $\epsilon_\nu^\mathrm{fast} = Q T_9^6$ erg cm$^{-3}$
s$^{-1}$.  The simplest and most efficient fast process is the direct
Urca (``DUrca'') with nucleons and this process has $Q \simeq 10^{27}$.
In the presence of hyperons other DUrca processes are possible with
slightly reduced efficiencies.  Other fast, but less efficient,
processes are possible in the presence of a meson condensate (either
pion or kaon) that also have a $T^6$ dependence and $Q$ in the range
$10^{24} - 10^{26}$ for pions and $10^{23} - 10^{25}$ for kaons.
Finally, if the neutron star has an inner core with deconfined quark
matter, similar DUrca processes are present with efficiencies that can
match the nucleon DUrca one.

The resulting neutrino luminosity is then given by
\be
L_\nu^\mathrm{slow} \approx \frac{4}{3} \pi R^3 \cdot Q^\mathrm{slow} T_9^8 \equiv N^\mathrm{slow} T_9^8
\label{Eq:Lnu_slow}
\ee
and
\be
L_\nu^\mathrm{fast} = \frac{4}{3} \pi R_p^3 \cdot Q^\mathrm{fast} T_9^6 \equiv N^\mathrm{fast} T_9^6
\label{Eq:Lnu_fast}
\ee
where $R$ is the radius of the neutron star and $R_p$ the radius of the inner core (i.e., high density) region where the given fast process
is acting.
There are theoretical uncertainties on $Q$ of a factor of a few for the slow processes and the DUrca ones,
but they are much larger in the case of the meson condensates.
Moreover, for the slow processes $R \sim 10$ km while for the fast ones $R_p$ can range from $\sim 0$ km, i.e.
almost no fast neutrino emission, to almost 10 km depending on the equation of state (in the case it allows such processes)
and the mass of the neutron star.
We take as a typical value $R_p \sim 5$ km but varying it can change $N^\mathrm{fast}$ by almost three orders of magnitude.

In the presence of pairing (causing superfluidity and/or superconductivity), neutrino emission processes can be strongly altered
(see, e.g., \citealt{2006NuPhA.777..497P}).
In low mass neutron stars where the MUrca processes can be strongly suppressed by pairing,
bremsstrahlung processes may become the dominant sources of neutrinos.
In the fast cooling scenarios $L_\nu^\mathrm{fast}$ can also be significantly reduced by pairing.
Moreover, pairing opens a new neutrino emission channel from the constant formation and breaking of Cooper pairs
(dubbed the ``pair breaking and formation'' or ``PBF'' process). The corresponding emissivity can be roughly approximated by 
$\epsilon_\nu^\mathrm{PBF} \approx Q T_9^8$ erg cm$^{-3}$ s$^{-1}$
where $Q$ can reach $10^{22}$ in optimal conditions depending on the type of pairing and its corresponding critical 
temperature $T_c$ \citep{2009ApJ...707.1131P}.

The photon luminosity is simply $L_\gamma = 4\pi R_\infty^2
\sigma_{SB} (T_e^\infty)^4$ where $\sigma_{SB}$ is the
Stefan-Boltzmann constant and $T_e^\infty$ the star's red-shifted
effective temperature.  $T_e$ has to be related to the internal
temperature $T$ and, as a simple rule, one can use $T_e \approx 10^6
T_8^{1/2}$ K.  This implies that $L_\gamma \simeq 7\times
10^{32}T_8^2$ erg s$^{-1}$.  A detailed study of accreted neutron star
envelopes and the resulting $T_e - T$ relationship can be found in
\cite{1997A&A...323..415P} and \cite{2004A&A...417..169Y}.  The latter
work shows that the $T$ dependence of $L_\gamma$ range between
$T^{1.7}$ to $T^{2.3}$, depending on the actual chemical composition
of the accreted envelope.

\subsection{Comparison of deep crustal heating with data}
\label{Sec:crustheating_data}

\begin{figure}
 \begin{center}
\includegraphics[width=0.99\columnwidth]{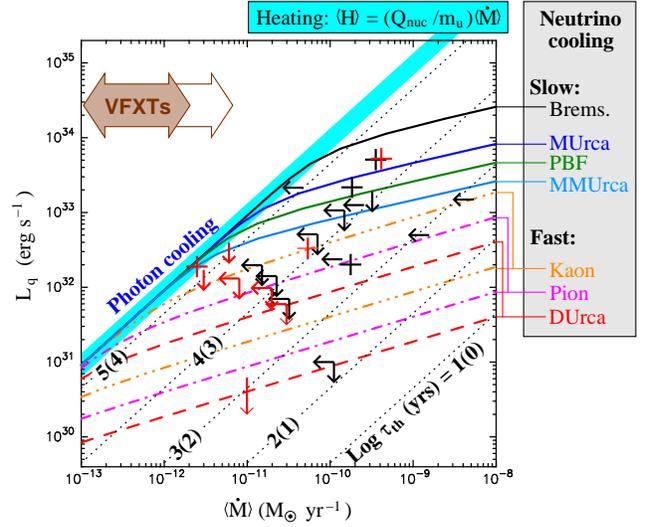}
    \end{center}
\caption{Locations of steady state quiescent luminosity $L_q$ versus averaged mass acretion rate, $\langle \dot{M}\rangle$
for a series of cooling scenarios. 
Each curve, for the various labeled neutrino cooling scenarios, is given by the energy balance of Eq.~(\ref{Eq:quiescent})
and the ``Heating'' band show the predicted range of $\langle H \rangle$ values:
(see Section~\ref{Sec:crustheating_data} for details).
Dotted lines are tracks of constant $\tau_{th}$, from Eq.~(\ref{Eq:tau}), for high specific heat
and, in parenthesis, low specific heat (i.e, $C_V = 10^{30}\, T$ erg K$^{-1}$ and $10^{29}\, T$ erg K$^{-1}$, 
or $f_{SF} = 1$ and $0.1$, respectively).
The displayed observational data are taken from \protect\cite{2010ApJ...714..894H}.
}
 \label{Fig1}
\end{figure}

Predictions for $L_q$ as a function of $\langle \dot{M} \rangle$ for
the various neutrino emission scenarios described above are shown in
Fig.~\ref{Fig1}, and compared with data (similar to what has been done
by
\citealt{2004ARA&A..42..169Y,2007ApJ...660.1424H,2009ApJ...691.1035H,2010ApJ...714..894H}).

In this figure the band ``Heating'' shows the average heating rate $\langle H \rangle$
from Eq.~(\ref{Eq:Heat}) with $Q_{nuc}$ ranging from 1 to 2 MeV.
Any star located on this line balances its heating only by its $L_q$
and is thus in the {\em photon cooling regime} of Eq.~(\ref{Eq:LqMdot}).
However, an object located below it has significant neutrino losses: 
it is in the {\em neutrino cooling regime} and the difference between its
 observed $L_q$ and the corresponding value of $\langle H \rangle$,
at the same \averagedmdot, on the ``Heating'' lines directly gives its $L_\nu$
from Eq.~(\ref{Eq:quiescent}).

Each curve in Fig.~\ref{Fig1}, for the various neutrino cooling scenarios, is given by
the energy balance of Eq.~(\ref{Eq:quiescent}).  
The parameters set we use is consistent with the
one proposed by \cite{2003A&A...407..265Y}.
The photon luminosity
is choosen as $L_\gamma = 7 \times 10^{34} \, T_9^2 \,\, \lum$ and the
heating rate, or luminosity, is obtained from Eq.~(\ref{Eq:Heat}) with
$Q_{nuc} = 1.5$ MeV. The four upper lines, ``Brems.'', ``MUrca'',
``PBF'', and ``MMUrca'' correspond to the four slow neutrino cooling
scenarios.  We use, for the correspondings $L_\nu$ in
Eq.~(\ref{Eq:Lnu_slow}), $N^\mathrm{Brems}=5\times 10^{37}$,
$N^\mathrm{MUrca}=5\times 10^{39}$, $N^\mathrm{PBF}=5\times 10^{40}$,
and $N^\mathrm{MMUrca}=5\times 10^{41}$.  As mentioned above, the
``PBF'' scenario can have a very wide range of efficiencies and the
case considered here corresponds to the most efficient one as deduced
for maximal compatibility of the ``minimal cooling'' scenario with
data from isolated cooling neutron stars
\citep{2004ApJS..155..623P,2009ApJ...707.1131P}, and with the
interpretation of the observed cooling of the neutron star in the
supernova remnant Cassiopeia A
\citep{2011PhRvL.106h1101P,2012arXiv1206.5011P}.
The ``MMUrca'' scenario actually covers a wide range of neutrino emission 
efficiencies that strongly depend on the neutron star mass and smoothly 
merge into the pion cooling scenario.

The three pairs of lines ``Kaon'', ``Pion'', and ``DUrca'' show the prediction for each
corresponding scenario when maximum and strongly reduced $L_\nu^\mathrm{fast}$
are assumed. We use, for the
correspondings $L_\nu$ in Eq.~(\ref{Eq:Lnu_fast}), $N^\mathrm{Kaon}$
ranging from $5\times 10^{40}$ to $5\times 10^{43}$, $N^\mathrm{Pion}$
from $5\times 10^{41}$ to $5\times 10^{44}$, and $N^\mathrm{Durca}$
from $5\times 10^{41}$ to $5\times 10^{45}$. When $L_\nu \gg L_\gamma$
one has $L_\nu^\mathrm{fast} = \langle H \rangle$ and since
$L_\nu^\mathrm{fast} \propto T^6$, if the neutrino efficiency is
reduced by a factor $10^3$, $T^2$ must be 10 times higher for
$L_\nu^\mathrm{fast}$ to keep matching $\langle H \rangle$.  Since
$L_\gamma \simprop T^2$, the resulting predicted $L_q$ is an order of
magnitude higher.  
Reduced efficiency of $L_\nu^\mathrm{fast}$ can be
either due to an emissivity lower than quoted above or due to a
smaller $R_p$.  
Thus, to each one of these fast neutrino processes corresponds
a band of at least one order of magnitude width in predicted $L_q$.

Comparison of these predictions with the data first shows that only a
few objects are on the ``photon cooling'' line, i.e., are described by
Eq.~(\ref{Eq:LqMdot}), and that a large number of the observed quiescent neutron-star
LMXBs have an inferred $L_q$ which requires significant neutrino
emission \citep[e.g.,][]{2001ApJ...548L.175C}.  In most case one sees
that $L_\nu$ dominates over $L_\gamma$ by one to two orders of
magnitude.  Enhanced neutrino emission implies a large neutron star
mass that must have been accreted during the lifetime of the X-ray
binary, unless most neutron stars can be born this massive.  The
latter possibility would conflict with the observation of young
isolated cooling neutron stars: the cooling of these stars is driven
by neutrino emission and their observed luminosity can be reproduced
with a $L_\nu$ similar to the ``PBF'' model of Fig.~\ref{Fig1} (see,
e.g., \citealt{2009ApJ...707.1131P})\footnote{This curve corresponds
to the minimal cooling scenario, but any scenario fitting the isolated
neutron stars data must have a similar $L_\nu$.}.

The conclusion is that, unless neutron stars in LMXB are born more
massive that isolated ones, a large fraction of the neutron stars shown in
Fig.~\ref{Fig1} must have accreted a significant amount of mass during
their lifetime.  A way around this conclusion is that the presently
deduced $\langle \dot{M} \rangle$ is not representative of the recent
history of these systems (i.e., \averagedmdot~was significantly lower
in the past) and for this we examine in the next subsection their
thermal inertia.  An alternative is that there is something missing in
the deep crustal heating model.

\subsection{Thermal response time-scale}
\label{Sec:tauth}

\begin{figure}
 \begin{center}
\includegraphics[width=0.9\columnwidth]{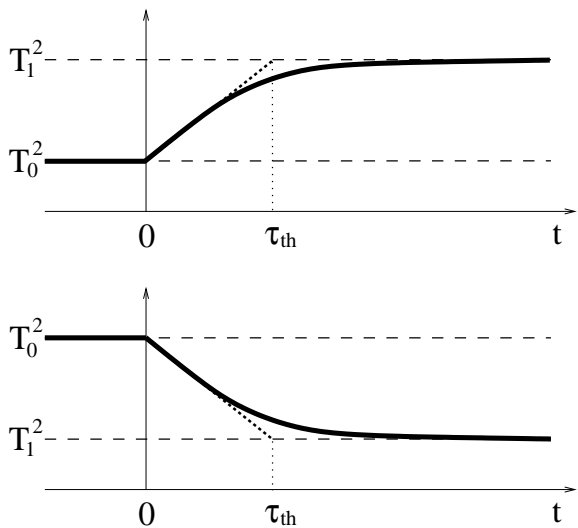}
    \end{center}
\caption{Schematic evolution of star's core $T^2$ after an abrupt increase (top panel) or
decrease (lower panel) in $\langle \dot{M} \rangle$. 
See Section~\ref{Sec:tauth} for details.}
 \label{Fig2}
\end{figure}

Since human X-ray observations of neutron-star LMXBs only span, in the best case, a few decades,
the estimated  $\langle \dot{M} \rangle$ are highly uncertain in many cases.
Of particular importance is an estimate of the thermal response time-scale $\tau_\mathrm{th}$
of a neutron star after a significant change of $\langle \dot{M} \rangle$.
This can be easily obtained from Eq.~(\ref{Eq:cool}) after specifying $C_V$.
Most of a neutron star specific heat is provided by its core and, for degenerate Fermions, $C_V \propto T$
so we simply write $C_V = C \cdot T$ erg K$^{-1}$ \citep{2004ARA&A..42..169Y,2006NuPhA.777..497P}.
Then Eq.~(\ref{Eq:cool}) gives
\be
\frac{d(T^2)}{dt} = \frac{2}{C} (\overline{H} - \overline{L})
\label{Eq:cool2}
\ee
where $\overline{H}$ and $\overline{L}$ are short term time averages of the heating term $H$ and
the cooling term $L = L_\gamma + L_\nu$.
By ``short term'' time average we mean averaged over many accretion outbursts but during a
time span still much shorted that $\tau_{th}$.
We assume that at time $t=0$ there is an abrupt change of $\overline{H}$ from $\overline{H}_0$ to $\overline{H}_1$,
and that at times $t < 0$ the star was in a stationary state at a temperature $T_0$.
Then at $t<0$ one had $\overline{L}_0 = \overline{H}_0$.
At $t>0$ the star begins to react and the resulting evolution of $T^2$ is illustrated by the thick curves in Fig.~\ref{Fig2}.
At times runs $T$ will tend toward a new stationary state: $T \rightarrow T_1$ and 
$\overline{L} \rightarrow \overline{L}_1 = \overline{H}_1$.
One can estimate the thermal response time $\tau_{th}$ by using a straight line (dotted line in the figure)
from the initial position at $t=0$ and $T=T_0$ till it reaches $T_1$ at time $\tau_{th}$.
The slope of this $T^2$ trajectory at $t=0$ is simply $2(\overline{H}_1 - \overline{L}_0)/C$
since $\overline{L} = \overline{L}_0$ has not yet evolved.
This gives us
\begin{eqnarray}
\tau_{th} \approx \frac{C \, (T_1^2 - T_0^2)}{2 \, (\overline{H}_1 - \overline{L}_0)} =
\frac{C \, (T_1^2 - T_0^2)}{2 \, (\overline{H}_1 - \overline{H}_0)}
\nonumber
\end{eqnarray}
were we also used that $ \overline{L}_0 =  \overline{H}_0$.
If $\overline{H}_1 \ll \overline{H}_0$ or $\overline{H}_1 \gg \overline{H}_0$
this expression for $\tau_{th}$ will be dominated by the high $\overline{H}$, high $T$, term and we can write
\be
\tau_{th} \approx \frac{CT_\mathrm{high}^2}{2 \langle H \rangle_\mathrm{high}}
\label{Eq:tau}
\ee
Models show that $C \approx 10^{30}$ when $T$ is measured in Kelvin, with only a small dependence on the mass of the neutron star.
However it can be reduced by up to a factor 10 in the presence of pairing \citep{2006NuPhA.777..497P},
i.e., $C \rightarrow f_{SF} C$, with $f_{SF}$ ranging from 1 (no pairing) down to 0.1 (maximum extend of
superfluidity/superconductivity).
$\langle H \rangle$ is obtained from $\langle \dot{M} \rangle$ with Eq.~(\ref{Eq:Heat}).
Notice that since $L_\gamma \simprop T^2$ one has the fortuitous, but useful, result
\be
\tau_{th} \approx 2 \times 10^4 \, f_{SF } \, \frac{L_{q}}{10^{32} \lum} \, \frac{10^{-11} \mdot }{\langle \dot{M}\rangle} \; \mathrm{yrs} \, .
\label{Eq:tau2}
\ee
Several tracks (the dotted lines) for various values of $\tau_{th}$ are also shown in Fig.~\ref{Fig1}.

Notice that the obtained $\tau_{th}$ is just the Kelvin-Helmoltz time-scale $\tau_{KH} = E_{th}/L$ since
$E_{th} = \int C_v dT = \frac{1}{2} CT^2$.
However, when a star transit between a high and a low $\langle \dot{M}\rangle$ states we have two $\tau_{KH}$,
and our analysis tells us that the relevant one is the high state one.
If the energy loss is due to neutrinos the high state $\tau_{KH}$ is the shortest one since $L_\nu \propto T^8$ or $T^6$
and thus $\tau_{KH} \propto T^{-6}$ or $T^{-4}$.
This is intuitively clear from Fig.~\ref{Fig2}: 
if the star starts from a low $\langle \dot{M}\rangle$ its {\em heating} is driven by the new high $\langle \dot{M}\rangle$
while if the star start from a high $\langle \dot{M}\rangle$ its {\em cooling} is driven by the remaining previous high $L_\nu$.
However, if the cooling is driven by the photon luminosity $L_\gamma \propto T^2$ then $\tau_{th}$ is essentially independent
of $\langle \dot{M}\rangle$:
Eq~(\ref{Eq:Heat}) and (\ref{Eq:tau2}) give $\tau_{th}^{\gamma} \approx 10^5 f_{SF}$ yrs,
as is confirmed by the $\tau_{th}$ line in Fig.~\ref{Fig1}.
In both cases, it is simply the shortest $\tau_{KH}$ that determines the time-scale.

Returning to the issue raised at the end of
Section~\ref{Sec:crustheating_data} about the mass of these neutron
stars, we see from Fig.~\ref{Fig1} that the ones requiring fast
neutrino emission, and hence a high mass, have a $\tau_{th} \sim 10^3
- 10^4$ yrs. Postulating that they have a low $L_q$ because $\langle
\dot{M} \rangle$ was much lower in the past and suddenly increased
recently would require that this increase occurred in the last
$10^3-10^4$ for all of them.  Such a claim is hardly sustainable and
the most natural interpretation of the large spread in observed $L_q$,
for a given $\langle \dot{M} \rangle$, is the result of a large range
of neutron star masses: more massive stars should have a larger
neutrino emission efficiency and hence appear fainter.
 
\subsection{Evolution of neutron stars with varying $\dot{M}$}
\label{Sec:crustheating_evolve}

When $\langle \dot{M} \rangle$ changes, the $L_q$ of the neutron star will slowly evolve on a time-scale $\tau_{th}$.
It will follow a trajectory, in the $L_q$ vs $\langle \dot{M} \rangle$ plane, given by Eq~(\ref{Eq:quiescent})
which is just one of the many curves plotted in Fig.~\ref{Fig1}, depending on its neutrino emission efficiency (i.e., and therefore very likely its mass).

If neutrino cooling dominates the energy balance, i.e., $L_\nu = \langle H \rangle$, we have for the star's core temperature $T$:
$T^2 \propto \langle H \rangle^{1/3}$ for fast ($T^6$) neutrinos and $T^2 \propto \langle H \rangle^{1/4}$ for
slow ($T^8$) neutrinos. 
Since $L_q \simprop T^2$ we obtain for fast neutrino emission
\be
L_q \propto \langle \dot{M} \rangle^{1/3}
\label{Eq:LM_fast}
\ee
and
\be
L_q \propto \langle \dot{M} \rangle^{1/4}
\label{Eq:LM_slow}
\ee
for slow neutrino emission.
For low enough $\langle \dot{M} \rangle$ these neutrino trajectories curve down and merge into the photon cooling trajectory
$L_q = (Q_{nuc}/m_u) \langle \dot{M} \rangle$.
These simple power-law behaviors are clearly seen in Fig.~\ref{Fig1}.
However, for a very rapid change in $\dot{M}$ the neutron star may be found ``off-trajectory''
for a time $\sim \tau_{th}$ that can be as large as $10^5$ years for very faint systems.

Notice that from the theory point of view the photon trajectory is the cleanest one:
it only depends on $Q_{nuc}$, Eq.~(\ref{Eq:LqMdot}),
which is known within a factor two, 1 - 2 MeV, within the deep crustal heating model.
The neutrino trajectories (which depend on the core temperature $T$) are plagued with the uncertainty on
how to relate $T_e$ with the core $T$.
This $T_e - T$ relationship depends on the outer layers chemical composition and the star's surface gravity:
for a given neutron star model (encapsulated in our parameters $N$ and $C$ for the neutrino luminosity and the specific heat)
the predicted $L_q$ can vary by almost one order of magnitude because of the $T_e - T$ relationship
\citep{1997A&A...323..415P,2004A&A...417..169Y}.

\subsection{Testing the deep crustal heating model}

So far this deep-crustal heating model has mostly been tested using
sources within a relatively small $\langle \dot{M} \rangle$ range,
between $5 \times 10^{-12}$ to $5\times 10^{-10}$ M$_\odot$ yr$^{-1}$,
as seen from Fig.~\ref{Fig1}.
For higher $\langle
\dot{M} \rangle$ the model will be difficult to test because it
straddles in the range of the persistently bright sources and
quiescent measurements are impossible for such sources. However, at
the low $\langle \dot{M} \rangle$, between $1 \times 10^{-13}$ erg
s$^{-1}$ and $5 \times 10^{12}$ erg s$^{-1}$, it is likely that
sources can be added; in particular the neutron-star VFXTs can have
$\langle \dot{M} \rangle$ in this range (see
section~\ref{mdotestimate}). We now discuss the prospect of
using those systems to test the deep-crustal heating model \citep[see
also the brief discussion in][]{2008AIPC.1010..382W}.

\section{The quiescent luminosity of neutron-star VFXTs} \label{Sec:qLx}

From the results of the previous section, illustrated in
Fig.~\ref{Fig1}, we can estimate the expected quiescent luminosities
$L_q$ of VFXTs from estimates of their long term average accretion
rates.  However, it is notoriously difficult to estimate $\langle
\dot{M} \rangle$ for any X-ray transient and even more so for VFXTs
because they are difficult \citep[despite the sensitivity of
X-ray instruments in orbit;][]{2006A&A...449.1117W} to detect and
therefore most of their outbursts are likely missed.  For the
deep-crustal heating model, the \averagedmdot~which is needed is the
$\dot{M}$~averaged over a time $\tau_{th}$ that (as can be seen 
from Fig.~\ref{Fig1}) ranges between $10^4$ and $10^5$ years at
$\dot{M} \sim 10^{-12} \, \mdot$ and $L_q \sim 10^{31} \, \lum$.


\subsection{Estimation of $\mathbf{ \langle \dot{M} \rangle}$ \label{mdotestimate} in VXFTs}

An estimation of an upper limit on \averagedmdot~for the VFXTs can be
obtained using some simple assumptions.  As already stated in the
introduction, VFXTs have peak X-ray luminosities (2--10 keV) $ < 1
\times 10^{36}$ erg s$^{-1}$. This limit was chosen in the
classification of \cite{2006A&A...449.1117W} because it roughly
corresponds to the sensitivity limit of the past and present X-ray
all-sky monitors in orbit (e.g., {\it BeppoSAX}/WFC, {\it RXTE}/ASM,
{\it Swift}/BAT, {\it Integral}, {\it MAXI}) for sources at 8
kpc. Typically those instruments are sensitive to X-ray outbursts
which have a peak flux (2--10 keV) above 10 mCrab\footnote{Some
instruments or programs are more sensitive \citep[e.g., the {\it
RXTE}/PCA bulge scan project reached about 1
mCrab;][]{2001ASPC..251...94S}. However, for the purpose of this
paper, we assume a conservative limit of 10 mCrab.}, which corresponds
to a 2--10 keV flux of $\sim 2 \times 10^{-10}$ erg s$^{-1}$ cm$^{-2}$
and a limiting 2--10 keV luminosity sensitivity of $1 \times 10^{36}$
erg s$^{-1}$ (for 8 kpc; which was assumed the typical distance towards
VFXTs since most have been found in the Galactic bulge).

Typically the bolometric luminosity is then a factor 2-3 higher
\citep[see][]{2007A&A...465..953I} and thus the upper limit on the
bolometric peak luminosity would be $< 3 \times 10^{36}$ erg
s$^{-1}$. Assuming perfect efficiency of the accretion process, this
accretion luminosity is related to the accretion rate in outburst
$\dot{M}$ through $L = {G M \dot{M} \over R}$, with $G$ the
gravitational constant, $M$ the mass of the neutron star and $R$ the
radius of the star. Here we use a ``canonical'' neutron star with a mass
of 1.4 solar masses and a radius of 10 km. The luminosity upper limit
then result in an upper limit on the peak mass accretion rate during
outburst of $<3 \times 10^{-10}$ M$_\odot$ yr$^{-1}$. Assuming that
the duty cycle DC (with DC = $t_o/ (t_o + t_q)$, $t_o$ being the
outburst duration time and $t_q$ the quiescent duration time) of VFXTs
is between 1\%-10\% (as is typically observed for the recurrent bright
transients), then \averagedmdot~$\lesssim 10^{-12} - 10^{-11}$
M$_\odot$ yr$^{-1}$. This is a rather conservative upper limit because
it assumes a ``step-function'' outburst in which the sources are always
accreting just below the limits set by the all-sky instruments when
they are in outburst. However many systems will have peak luminosities
well below this level \cite[e.g.,][]{2010A&A...524A..69D} and during
the outbursts the peak luminosities are only reached during a small
fraction of the outburst, similar to the outburst profiles of brighter
transients \citep[e.g., see][]{1997ApJ...491..312C}. This will
significantly lower this upper limit for the VFXTs. However, this
limit already shows that indeed VFXTs have very low \averagedmdot;
i.e., in the range which has hardly been used to test the deep-crustal
heating model.

Moreover, for several VFXTs more stringent constraints can be obtained
on their \averagedmdot. \cite{2010A&A...524A..69D} have estimated the
\averagedmdot~ of several VFXTs near the Galactic center using 4 years
of {\it Swift}/XRT monitoring data on Sgr A*. The majority of the
sources considered in that paper had \averagedmdot~$<2\times 10^{-12}$
M$_\odot$ yr$^{-1}$ and were typically in the range of $10^{-13}$ -
$10^{-12}$ M$_\odot$ yr$^{-1}$. This demonstrates that at least some
VFXTs have extremely low \averagedmdot~and it might be possible that a
large fraction of the VFXTs have similar \averagedmdot.

\subsection{Estimation of $L_q$ in VFXTs
\label{enhanced} \label{qLxestimate}}

A naive estimation of the possible range of $L_q$'s for neutron-star
VFXTs would simply consider the various trajectories of
Fig.~\ref{Fig1}: we can consider the sample of objects depicted in
Fig.~\ref{Fig1} and move them along their respective cooling
trajectories into the range of the (very low) \averagedmdot~above
estimated for the VFXTs.  There are, however, two immediate issues
with this extrapolation:

\begin{itemize}

 \item[1)] How representative are the above estimated
\averagedmdot~of VFXTs with their real long term \averagedmdot? \\

 \item[2)] Since the spread in the $L_q$ observed for the brighter transients
exhibited in Fig.~\ref{Fig1} is likely due to a large spread in
neutron star masses (as argued above) how representative is the mass
distribution of the bright transients for the mass range of the VFXTs?

\end{itemize}

\subsubsection{The \averagedmdot~versus $\tau_{th}$ issue}

The above limits on the \averagedmdot\, assumes that the current
observed behavior of VFXTs is representative of their general
behavior.  However, this does not need to be true, and it might be
that some fraction of the VFXTs are actually sources which usually
accrete at much higher rates and we only observe them during a small
period in which their \averagedmdot\, is much less \citep[although see
the discussions in][demonstrating that this is likely not true for the
majority of the VFXTs]{2006MNRAS.366L..31K,2006A&A...449.1117W}.

The issue is how long has this period of very low \averagedmdot~lasted compared to
the neutron star relaxation time $\tau_{th}$.
As discussed in Section~\ref{Sec:crustheating_evolve}, if the time-scale of the evolution of
\averagedmdot~is larger than the corresponding $\tau_{th}$ the neutron star will evolve
along one of the trajectories exhibited in Fig.~\ref{Fig1}:
this is the Case B discussed below.
However, in the opposite case we have:

{\bf Case A:} 
The VFXTs in this class are normally bright transients
or even bright persistent sources and we only observe them during a
short-lived VFXT episode, shorter than their $\tau_{th}$.
This is not unfeasible, because several
bright systems have exhibited faint to very-faint outbursts as well
\citep[e.g.,][]{2007ATel.1055....1L,2009A&A...495..547D,2011ApJ...736..162F,2012arXiv1204.6043D}. 
In other words, the estimated \averagedmdot~is rather an $\overline{\dot{M}}$
in the notations of Section~\ref{Sec:tauth} and the star has not yet adjusted to the
new low $\overline{\dot{M}}$:
the star's present $L_{q}$ is still determined by its previous higher \averagedmdot~and
is ``off-trajectory''.
In the case the neutron star is not too massive and was previously on a
slow neutrino cooling track its present $L_q$ can be in the range of 
$10^{33-34} \, \lum$:
such high $L_q$ can be sustained for more than $10^5$ yrs, the corresponding $\tau_{th}$
of bright transients on the slow cooling trajectories
($\tau_{th}$ being determined by the value in the high \averagedmdot~regime).
With such an $L_q$ and an \averagedmdot~in the VFXT regime it would be located in the
diagram of Fig.~\ref{Fig1} above the photon cooling line. This is a region
that is unaccessible for a star in thermal equilibrium.

{\bf Case B:} 
The observed behavior of the VFXTs in this class is
representative for their behavior on a long enough time that they are in
a steady state, i.e., they can be correctly located on one of the trajectories of Fig.~\ref{Fig1}. However, for the Case B systems, one has to consider the second issue, which we will address in the next section.

\subsubsection{The mass distribution issue}

Extrapolating the observed distribution of $L_q$ of bright transients
to the low \averagedmdot~regime could allows us to predict the
expected distribution of $L_q$ for the VFXTs.  However, isolated young
neutron stars, that likely have a mass distribution between 1.2 to 1.6
$M_\odot$, with an observed thermal luminosity are well described by
slow neutrino emission processes
\citep{2004ApJS..155..623P,2009ApJ...707.1131P}.  This is in sharp
contrast with what is observed for the bright neutron-star
transients. Most of them have an $L_q$ that requires fast neutrino
processes, implying a mass distribution strongly skewed toward higher
masses due to the long term accretion.  As a consequence, the $L_q$
distribution of the VFXTs that have been in the VFXT phase for long
enough that they are in a steady state (Case B) should show a
significant imprint of their very long term past accretion history and
Case B can be divided into two distinct sub-cases:

{\bf Case B1}: Some systems might have been a VFXT for a period longer
than $\tau_{th}$ but before that they were similar to the brighter
transients. Although originally the temperature of the core should be
relatively high, the time spend in the VFXT phase is long enough that
the neutron star has adjusted its core temperature (i.e., it has
become colder) to the very low accretion rate, but during the time
spend before this phase (when the accretion rate was much higher) the
source could have accreted enough material for it to cross the mass
threshold beyond which fast core cooling could occur.  Therefore,
those systems should be on one of the ``fast neutrino cooling''
trajectories of Fig.~\ref{Fig1} and be significantly fainter in
quiescence than predicted using the standard deep-crustal heating
model, Eq.~(\ref{Eq:LqMdot}).

{\bf Case B2}: Some systems might have been VFXTs throughout their
life and the estimated \averagedmdot~is representative for the
\averagedmdot~throughout their life \citep[we call them primordial
VFXTs; see also][who talked about the possible existence of such
systems]{2006MNRAS.366L..31K}. Therefore, since it is expected that
LMXBs live for $10^{8-9}$ years, only about $10^{-5}$ to $10^{-2}$
solar masses (assuming \averagedmdot~$=10^{-13} - 10^{-11}$ M$_\odot$
yr$^{-1}$; Section~\ref{mdotestimate}) has been accreted by the
neutron stars. Therefore, it is quite likely that most, if not all, of
these sources have not been accreting enough matter during their life
to increase their masses (assuming they are all born with a mass of
$\sim$1.4 M$_\odot$) above the threshold so that enhanced core cooling
would become active.  If true these systems should be on the slow
neutrino cooling tracks of Fig.~\ref{Fig1} that, in the estimated
\averagedmdot~range of VFXTs, merge with the photon cooling track.
Therefore, they should agglutinate on the narrow ``photon cooling''
region and indeed be detectable at $10^{31-33}$ erg s$^{-1}$.
However, this assumes that the standard deep-crustal heating is active
in those sources, but in the next section we argue that this might not
be the case.

\subsection{Non-standard heating in neutron-star VFXTs \label{hybrid}}

\begin{figure}
 \begin{center}
\includegraphics[width=0.9\columnwidth]{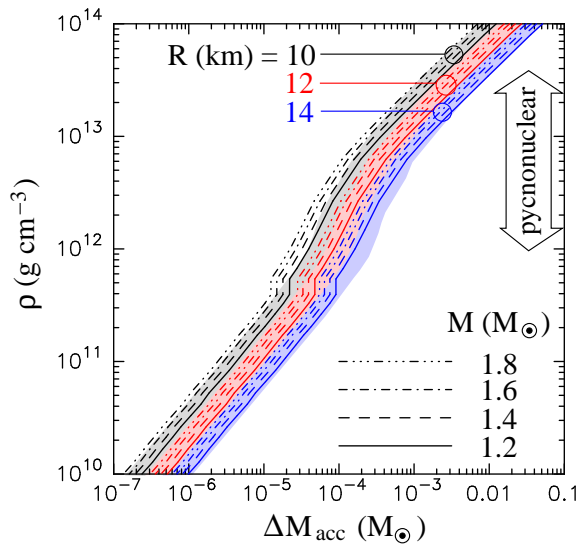}
    \end{center}
\caption{Estimate of the density reached by accreted matter as a function of total accreted mass $\Delta M_{acc}$
for 4 different neutron star masses and 3 different radii.
Lines show the values obtained assuming a catalyzed crust EOS while the background shaded areas show
the corresponding range of values for a wholly accreted crust.
The vertical ``pycnonuclear'' double-arrow marks the density range in which pycnonuclear fusions are expected to take place,
and significantly heat the neutron star, once catalyzed matter has been replaced by accreted matter. The jump in the diagrams at densities $\sim 5 \times 10^{11}$ g cm$^{-3}$ represents the onset of neutron drip.
See Section~\ref{hybrid} for details.}
 \label{Fig3}
\end{figure}

In the deep-crustal heating scenario, it is explicitly assumed that
the original, catalyzed crust is fully replaced by an accreted one
\citep{1990A&A...227..431H,2003A&A...404L..33H,2008A&A...480..459H}. The
low-Z elements present (at high densities) in an accreted crust
facilitate a significant amount of energy release due to the
pycnonuclear reactions. 
Those pycnonuclear reactions occur in the density range from
$10^{12}$ up to $\sim 5\times 10^{13}$ g cm$^{-3}$
\citep{1990A&A...227..431H,2003A&A...404L..33H,2008A&A...480..459H}. 
It is quite possible that among the
primordial VFXTs (the above Case B2 sources) a group of systems exist in which the crust is not
replaced to this depth and a partly accreted and a partly original
(albeit further compressed) crust (a hybrid crust) might be present.

To make this last assertion more quantitative, we show in Fig.~\ref{Fig3}
estimates of the amount of matter that needs to have be accreted in order
to replace the original catalyzed matter.
The needed accreted mass is simply
\be
\Delta M_{acc} =  4 \pi R^2 y
\label{Eq.Macc}
\ee
where $y$ is the accreted column density.  This column density
determines the pressure reached by the accreted matter since, from
hydrostatic equilibrium, $P=gy$ where $g = \mathrm{e}^{-\phi} \,
GM/R^2$ is the gravity acceleration in the crust and $\mathrm{e}^\phi
= \left(1-2GM/Rc^2 \right)^{-1/2}$ the red-shift ($G$ being the
gravitational constant and $c$ the speed of light).  For a given
pressure we can obtain the corresponding density $\rho$ using a crust
equation of state (EOS).  The EOS of a hybrid crust remains to be
calculated but we can bracket it between the EOS of a catalyzed crust
and a wholly accreted one.  For a catalyzed crust we use the EOSs of
\cite{1989A&A...222..353H} for the outer crust and
\cite{1973NuPhA.207..298N} for the inner crust.  For an accreted crust
we use the model of \cite{2008A&A...480..459H}.  The results of
Fig.~\ref{Fig3} show that, fortunately, the needed $\Delta M_{acc}$
does not depends strongly on the assumed EOS.  Pycnonuclear reactions
provide about 60-70\% of the total $Q_{nuc}$ and one sees that, for a
low mass ($1.2 \, M_\odot$) extended (14 km radius) neutron star,
about $2 \times 10^{-2} \, M_\odot$ of matter needs to be accreted so
that $Q_{nuc}$ reach its optimal value while about $2 \times 10^{-4}
\, M_\odot$ is needed for the accreted matter to reach the threshold
of the first possible reaction.  For a heavy ($1.8 \, M_\odot$)
compact (10 km radius) neutron star the corresponding numbers are
approximately 1 order of magnitude smaller but such a heavy star will
likely undergo fast neutrino cooling and have such a low $L_q$ in the
$\langle \dot{M} \rangle$ regime of primordial VFXTs that it will be practically
unobservable. We note that if such heavy
neutron stars could indeed be present in some primordial VFXTs that those neutron
stars have to have been born this massive.

The effect on the heating is not clear. It depends on to what depth
the crust has been replaced since pycnonuclear reactions occur down to
a density of $\sim 5\times 10^{13}$ g cm$^{-3}$ but can start already
at significantly lower density \citep[see,
e.g.,][]{2008PhRvC..77d5807H}. In particular most of the heat due to
pycnonuclear reactions is released in the density range of $10^{12}$
to $10^{13}$ g cm$^{-3}$ \citep[e.g.,][]{2008A&A...480..459H} and
significantly less mass has to be accreted to replace the crust to
those densities (Fig.~\ref{Fig3}). However, in the most extreme case in which the total
amount of accreted matter is only $10^{-5}$ M$_\odot$ then only the
outer crust, with densities below the neutron drip, has been fully
replaced, significantly inhibiting pycnonuclear reactions to occur in
the inner crust. Therefore, there might be a sub-group among the
primordial VFXTs for which the heating is significantly reduced which
would make them even fainter than already inferred from their low
\averagedmdot. However, other than fusion reactions, substantial heating may occur just below the neutron drip via cascades of electron capture and neutron emissions \citep[][]{2008PhRvL.101w1101G}. Therefore, detailed calculations are required to fully grasp the effect of a partially-accreted crust on the thermal state of transiently accreting neutron stars.

\section{Discussion} \label{sec:discussion}

We have estimated the quiescent thermal luminosity of neutron-star
VFXTs in order to determine if they can be used to test the
deep-crustal heating model in an hardly explored
\averagedmdot~regime. Unfortunately a conclusive answer cannot be give
due to the large uncertainties in our knowledge of the accretion rate
history of VFXTs. The \averagedmdot~of the source during the last
several thousands to tens of thousand years determines how much heat
has be deposited in the neutron star over that period and therefore
the thermal state of the star. However the long-term history over the
lifetime of the binary determines the amount of matter accreted and
therefore if enough matter has been accreted to trigger enhanced
neutrino emission processes in the core and if enough matter is
accreted to allow the activation of all pycnonuclear heating reactions
in the inner crust.

This last point arises because it is well possible that the amount of
matter which primordial VFXTs have accreted during their lifetime is
not enough to fully replace the original crust, leaving a crust which
is partly replaced by accreted matter and partly still contains the
original, albeit further compressed material. It is unclear how such a
hybrid crust would react to the accretion of matter and how this would
effect the thermal state of the neutron star. Likely less heat is
produced because not all pycnonuclear reactions can occur, but it is
not clear if other properties of a hybrid crust are also significantly
different compared to a fully accreted crusts, such as the thermal
conductivity. Besides obtaining more observational data to constrain
the models, detailed theoretical calculations have to be performed to
investigate the heating and cooling in neutron stars which have a
hybrid crust. In particular it is important to investigate different
crustal compositions with a variety of amount of matter accreted
\cite[e.g., an update study of the one performed
by][]{1979PThPh..62..957S}. This problem might not only be interesting
for VFXTs, but also to other types of neutron stars because VFXTs
might not be the only sources which harbor neutron stars with hybrid
crusts (see Section~\ref{additional}). Furthermore, VFXTs might be
important to understand low-luminosity X-ray sources in globular
clusters (Section~\ref{globulars}). In addition, they might form
an interesting group of sources to try to study cooling of the neutron
star crust (Section~\ref{forward}) after it has been heated during
outbursts.

\subsection{Additional potential sources without fully accreted crusts \label{additional}}

Despite that it is generally accepted that most neutron-star LMXBs are
rather old systems with ages of $10^{8-9}$ year, there are individual
sources which likely are much younger.  One relatively young system
might be the recently discovered transiently accreting  11 Hz X-ray
pulsar IGR J17480--2446 in the globular cluster Terzan 5
\citep[also called Terzan 5 X-2;][]{2010ATel.2929....1S,2011A&A...526L...3P}. 
This system is an
unusual LMXB because it was expected that the neutron stars in LMXBs
should have spin periods $<10$ millisecond because they are spun up by
the accretion of matter \citep[see review
by][]{1991PhR...203....1B}. The slow spin period of IGR J17480--2446 is
enigmatic and it has been hypothesized that this is due to the fact
that the system has so far only spend a relatively brief time in the
Roche-lobe overflow phase \citep[$10^7$ to $10^8$
years;][]{2012ApJ...752...33P}. If true, this system might be an
example of systems which do not have a fully replaced neutron star
crust.

The mass accretion rate of this source during outburst has been
estimated to be $3 \times 10^{-9}$ M$_\odot$ yr$^{-1}$
\citep{2011MNRAS.412L..68D}. The duty cycle of this system is poorly
constrained but if we assume again values of 1\%-10\% we obtain a
time-averaged accretion rate of $3 \times 10^{-11}$ - $3\times
10^{-10} $ M$_\odot$ yr$^{-1}$. Combined with the expected age of the
accretion phase this results in a mass accreted on the neutron star of
$3\times 10^{-4}$ to $3\times 10^{-2}$ M$_\odot$. Although the maximum
amount of matter accreted would indicate that the full crust is
replaced, it is also quite possible that the neutron star in this
system has a hybrid crust as well. \cite{2011MNRAS.412L..68D} found
the quiescent counterpart for this source to be rather cold,
significantly colder than expected using standard heating and cooling
theory. They suggested that in its neutron star enhanced core cooling
processes might be active although, as also shown above, probably not
enough matter has accreted on the star for the star to have become
massive enough to allow such processes to occur in the
core. Alternatively, they suggested that the duty cycle might be
extremely low, of the order 0.1\%.

Although not impossible, this duty cycle seems very low (and possibly
improbable in the disk instability model) and therefore we suggest an
another possible reason why the source is so faint in
quiescence\footnote{We note that the thermal quiescent luminosity of
the source is still well within the range observed from other
quiescent neutron-star LMXBs which would suggest that the source is
not special. This could indicate that the same physical processes are
at work in this source as well as in the other sources. This would
also satisfy the principle of Occam's razor, by not having to have to
postulate several mechanisms why certain quiescent LMXBs are colder
than expected by the standard theory.}: due to the presence of a hybrid
crust, not all the heating reactions can occur in the crust and
therefore less heat has been deposited in the neutron star to heat it
up to the expected temperature as inferred from its \averagedmdot.
This conclusion still holds when also taking into account that before
the Roche-lob overflow phase a wind-accretion phase
occurred. \cite{2012ApJ...752...33P} estimated that the mass accretion
rate in that phase would at most be $10^{-13}- 10^{-12}$ M$_\odot$
yr$^{-1}$. This phase could have lasted $10^{7-8}$ year and thus at
most $10^{-6} - 10^{-4}$ M$_\odot$ could have been accreted.

If this is the correct explanation for why the neutron star in IGR
J17480--2446 is colder than expected, one has to wonder if a similar
argument might also hold for other systems which have found to be too
cold. For example, the neutron star in SAX J1808.4--3658 seems to be
extremely cold
\citep[][]{2002ApJ...575L..15C,2007ApJ...660.1424H,2009ApJ...691.1035H}. Its
\averagedmdot~has been estimated to be $\sim 10^{-11}$ M$_\odot$
yr$^{-1}$ \citep{2007ApJ...660.1424H} and if it has lived shorter than
$10^{8}$ years the neutron star should have a hybrid crust. However,
this system is an accreting millisecond pulsar with a spin period of
401 Hz \citep{1998Natur.394..344W}. This means that a significant
amount of matter has to have been accreted by the neutron star to spin
it up to this spin frequency. Typically, the calculations show that
at up to 0.1 M$_\odot$ \citep{1987IAUS..125..393V} is needed to
accomplish this \citep[see review
by][]{1991PhR...203....1B}. Therefore, in SAX J1808.4--3658 the
neutron-star crust will have been fully replaced, which strongly
indicates that in the past the accretion rate of this system was
considerably larger than its current inferred \averagedmdot. Another
source which might be relatively young is Circinus X-1. The age of
this system is not known, but it has been suggested to be rather young
\citep[of the order of $<10^{4-5}$ years; see the discussion
in][]{2004MNRAS.348..458C}. Despite that it can accrete on occasions
at very high accretions rates (up to $>10^{-8}$ M$_\odot$ yr$^{-1}$),
this age (if confirmed) is sufficiently low that very likely not the
complete crust has been replaced. If the source would go fully
quiescence, it would be very interesting to determine the quiescent
luminosity of the neutron star in this system.

\subsubsection{Neutron stars in high-mass X-ray binaries}

In high-mass X-ray binaries (HMXBs) the neutron star accretes either from the strong stellar wind
of the companion (e.g., a supergiant star) or from the decretion disk
of a B type star, which is typically observed to be of type B0-B2 in
Be/X-ray transients \citep[see review by][]{2011Ap&SS.332....1R}. Such
early-type B stars only live between 10 to 30 million years. Typically
in Be/X-ray transients the sources have outbursts with X-ray
luminosities of $10^{36-37}$ erg s$^{-1}$ (corresponding to an
outburst accretion rate of $10^{-10}$ to $10^{-9}$ M$_\odot$
yr$^{-1}$) when the neutron star moves through the decretion disk at
periastron passage.  For some sources this occurs once every orbital
period (resulting in periodic outbursts; called type-I outburst) but
other sources are only occasionally in outburst. Therefore, it is
unclear what fraction of the time the neutron star is actually
accreting, but when assuming again a duty cycle of 1\%-10\%, this
would result in a \averagedmdot~of $10^{-12}$ to $10^{-10}$ M$_\odot$
yr$^{-1}$ and a total amount of mass accretion throughout the life
time of the system (assuming the system was a Be/X-ray transient for
the whole life of the B star which might be a significantly
overestimation of the duration of this phase) of $10^{-5}$ to $3
\times 10^{-3}$ M$_\odot$. Thus, it is quite possible that also the
neutron stars in some Be/X-ray transients have a hybrid crust. We note
that some systems also exhibited so-called type-II outbursts which are
much brighter (peak luminosities of $10^{38}$ erg s$^{-1}$) which can
last for weeks to months but they are very infrequent and not all
systems exhibit them. Therefore, we do not expect that those type of
outbursts will affect our main conclusion significantly.

Another class of HMXBs transients are the supergiant fast X-ray
transients \citep[or SFXTs; see, e.g.,][]{2011AdSpR..48...88S} in
which the neutron star transiently accretes from the variable dense
wind of a supergiant star. However, also very likely in those systems
the neutron star has only a partly replaced crust because the
supergiants only live very short and despite that the outbursts of
those systems can be very bright ($10^{38}$ erg s$^{-1}$) they are
very brief, very infrequent, and most of the time the neutron star
is only accreting at much lower rates or not at all. Although, if
before the supergiant phase the neutron star was also already
accreting significantly from the companion star \citep[e.g., during an
earlier Be phase;][]{2011MNRAS.415.3349L}, then more of the original
crust is replaced.

For the neutron stars in HMXB transients~\averagedmdot~is
typically higher than inferred for the VFXTs (they are typically more
in the range observed for the ordinary LMXB transients). Therefore, it is
expected that if standard heating and cooling occurs in those systems
\citep[as suggested by][]{1998ApJ...504L..95B}, that their thermal
emission should be readily detectable in quiescent. Enhanced core
cooling is not expected because they should be relatively light weight
neutron stars since little matter has been accreted \citep[although it
might be possible that some systems are born with massive neutron
stars; see, e.g.,][]{2001A&A...377..925B}. In contrast, the heating
might be affected by what type of crust is present (i.e., fully
accreted or hybrid crust) and HXMB transients might be very good
candidates to investigate the effect of hybrid crust on the thermal
properties of the neutron star. However, the situation for those
sources might be complicated by the much stronger magnetic field in
those systems ($10^{12-13}$ Gauss) compared to those of the neutron
stars in LMXBs ($10^{8-9}$ Gauss). It is unclear how strong the
effects of these stronger magnetic fields are on the heating and
cooling of the neutron stars and other related properties \citep[e.g.,
the thermal conductivity which is severally affects by super strong
magnetic fields of $>10^{13}$ Gauss and therefore likely also by
slightly lower
fields;][]{1999A&A...346..345P,2008A&A...486..255A}. More detailed
theoretical calculations have to be performed to determine the effect
of the magnetic field, in combination with the exact composition and
structure of the (possible hybrid) crust.

Observing HMXB transients in their quiescent state could be very
useful in this aspect. However, the number of neutron-star Be/X-ray
transients so far studied in quiescence is rather limited \citep[for a
source list see][]{2007ApJ...658..514R,2011ApJ...728...86T}. So far,
the obtained picture is complex. Some systems (like, e.g., EXO
2030+375) always remain rather bright in-between outbursts ($>10^{35}$
erg s$^{-1}$; basically they never transit to quiescence). However,
the majority of systems have quiescent luminosities between $10^{32}$
and $10^{34}$ erg s$^{-1}$
\citep[][]{2007ApJ...658..514R,2011ApJ...728...86T}. Spectral analysis
demonstrates that some systems are still very hard in quiescence with
power-law indices near 1 or even lower \citep[similar to what often is
seen in outburst;][e.g.,]{2007ApJ...658..514R}, while others are
softer with indices even up to 2.6
\citep[e.g.,][]{2002ApJ...580..389C}.

Although the quiescent data are usually not of very high quality,
several sources do not show pulsations in quiescence which might
indicate that indeed the accretion down to the surface has halted in
those systems
\citep[][]{2002ApJ...580..389C,2005ApJ...622.1024W}. However, in a few
other systems pulsations could still be detected in quiescence
demonstrating that in those systems either some of the matter still
reaches the neutron-star surfaces or the pulsations are in some way
caused by the interaction between the magnetic field (which is
rotating with the neutron star) and the accretion of matter down to
the magnetosphere \citep[][]{2007ApJ...658..514R,2005A&A...431..667M}.

For some systems it has been suggested \citep[][]{2002ApJ...580..389C}
that indeed the emission we observe is due to the cooling of the
neutron star and not due to some sort of accretion process, however,
the evidence is not conclusive due to the statistical quality of the
data. Furthermore, the possibility that they might harbor a neutron
star with a hybrid crust was not
considered. \cite{2011ApJ...728...86T} discussed possible reason why
the candidate Be/X-ray transient IGR J01363+6610 could not be detected
with {\it Chandra} in its quiescent state (e.g., the system containing
a black hole instead of a neutron star), but in light of the above
discussion we suggest that the possibility should be considered that
this source might still harbor a neutron star but one with a hybrid
crust which inhibits significant heating of the neutron star.

The situation for SFXTs is similar to that of the Be/X-ray transients
with only a handful of SFXTs studied in quiescence. Also those systems
show a variety in quiescent behavior \citep[see,
e.g.,][]{2005A&A...441L...1I,2010A&A...519A...6B,2012arXiv1207.3719B}. A
systematic and homogenous study of many more HMXB transients (both
SFXTs and Be/X-ray transients) in quiescence is needed to understand
fully how they can be used to study the deep-crustal heating model. A
survey (using {\it Chandra}) of 16 confirmed neutron-star Be/X-ray transients in their
quiescent state has recently been accepted (PI: Wijnands) which will
give more insight into this issue.

\subsection{VFXTs in globular clusters \label{globulars}}

Many faint X-ray sources have been found in the Galactic globular
clusters, and a large number are likely associated with neutron-star
X-ray transients \citep[see, e.g.,][]{1984MNRAS.210..899V}. But the
lack of a significant number of outbursts from those sources has lead
to suggestions that maybe those sources are associated with VFXTs
whose outbursts where missed by the all-sky monitors
\citep[][]{2008AIPC.1010..382W}. As estimated in
section~\ref{qLxestimate}, the quiescent X-ray luminosity of VFXTs in
the standard deep-crustal heating model would be in the range
$10^{31-33}$ erg s$^{-1}$ and indeed, if the standard heating and
cooling processes occur, a large fraction of the candidate quiescent
LMXBs could be associated with VFXTs. However, as also explained in
section~\ref{enhanced} and~\ref{hybrid} the quiescent luminosity of
VFXTs could be significantly lower than expected in the standard model
and therefore it is unclear if this conclusion still holds. Moreover,
even in the standard model it is unlikely that the VFXTs are
associated with the candidate quiescent LMXBs in globular clusters.

To demonstrate this, we rewrite the time-averaged accretion rate into

\begin{equation}
\langle \dot{M} \rangle = {\langle \dot{M}_o \rangle t_o + \langle \dot{M}_q \rangle t_q \over t_o + t_q} \approx   \langle \dot{M}_o \rangle {t_o  \over t_o + t_q} =  \langle \dot{M}_o \rangle~ {\rm DC} \label{DC}
\end{equation}

\noindent with $\langle \dot{M}_o \rangle$ the time-averaged accretion
rate in outburst and $\langle \dot{M}_q \rangle$ the time-averaged
accretion rate in quiescence. Equation \ref{DC} assumes that $\langle
\dot{M}_o \rangle t_o \gg \langle \dot{M}_q \rangle t_q$, which is
usually true but might not if $t_q \gg t_o$, thus for systems with a
extremely low duty cycle. Using equations~\ref{Eq:Heat} and \ref{DC}, and assuming $L_q = \langle H \rangle$ (thus standard, slow cooling) one
obtains

\begin{eqnarray}
DC = { L_q \over  10^{33} {\rm ~erg~ s}^{-1}} { 10^{-11}~ M_\odot ~{\rm
yr}^{-1} \over\langle \dot{M}_o \rangle } {  1.5 ~{\rm MeV}\over Q_{nuc}}\\
 >{ L_q \over 3 \times 10^{34} {\rm ~erg~ s}^{-1} } {  1.5 ~{\rm MeV}\over Q_{nuc}}\label{eqDC}
\end{eqnarray}

\noindent which assumes $ \langle \dot{M}_o \rangle <3 \times
10^{-10}$ $M_\odot$ yr$^{-1}$ (the limit set by the all-sky monitors).

Typically quiescent LMXB candidates in globular clusters have
bolometric X-ray luminosities of $10^{32-33}$ erg s$^{-1}$ with the
brightest being $\sim 3.6 \times 10^{33}$ erg s$^{-1}$ \citep[][about
a third of the sources listed in that paper have an X-ray luminosity
$>1 \times 10^{33}$ erg s$^{-1}$]{2003ApJ...598..501H}. This results
(using equation~\ref{eqDC}) in limits on the duty cycle of DC $>$
0.0025 - 0.005 for $L_q = 1 \times 10^{32}$ erg s$^{-1}$, DC $>$ 0.025
- 0.05 for $L_q = 1 \times 10^{33}$ erg s$^{-1}$ and DC $>$ 0.1 - 0.2
for $L_q = 3.6 \times 10^{33}$ erg s$^{-1}$. The range in DC is due to
the fact that we have assumed that $Q_{nuc} $ must be between 1 and
2. Putting reliable observational constraints on the duty cycle of
possible VFXTs in globular clusters is difficult. However, currently
there are about 30 quiescent LMXB candidates identified for which no
outbursts\footnote{This excludes the two sources (IGR J17480--2446,
Swift J174805.3--244637) in Terzan 5 which were previously identified
as candidate quiescent LMXBs \citep[][]{2006ApJ...651.1098H} but which
have now been shown to be associated with bright transients. However,
the final conclusions does not depend very sensitive on how many
quiescent LMXBs are currently known but have not been associated yet
with outbursts.}  have been seen yet (see \cite{2003ApJ...598..501H}
for a list, several additional sources have found since that
publication; e.g.,
\cite{2007ApJ...657..286L,2009MNRAS.392..665G,2011ApJ...738..129G,2012arXiv1208.1426M}). During
the {\it Chandra} and {\it XMM-Newton} observations which detected
those sources, they were not in outburst. Therefore, assuming that all
sources are from the same population of transient type, the duty cycle
of those systems is $<1/30 = 0.033$ \cite[see
also][]{2010AIPC.1314..135H}. We note that a number of clusters have
multiple {\it Chandra} and {\it XMM-Newton} observations during which
those sources were not seen in outbursts. However, those additional
observations do not constrain the duty cycle further because it is
quite possible that the quiescent duration is significantly longer
than the sampling time scale meaning that the observations probe the
same quiescent period and therefore are not statistically independent
(basically many outburst-quiescent cycles must have passed for the
sampling, if performed with random time-intervals between the
observations, to become independent). We note that the uncertainties
on the duty cycle inferred from observations is likely to be large,
but the duty cycle has to be $>$10\% to explain the brightest sources
(in the standard deep-crustal heating model) which seems
unlikely. Furthermore, similar low duty cycles were inferred for VFXTs
near the Galactic center \citep{2010A&A...524A..69D}, so it is quite
possible that they indeed have such low duty cycles.

The conclusion from the above exercise is that VFXTs can still be
associated with some of the quiescent LMXB candidates, but only with
the faintest sub-set of the group and likely not with the brightest
objects ($> 10^{33}$ erg s$^{-1}$; especially not those of a few times
$10^{33}$ erg s$^{-1}$; note that this assumed standard heating and
cooling). This is supported by the fact that the three VFXTs currently
known in globular clusters \citep[M15 X-3, NGC 6440 X-2, and NGC
6388;][]{2009ApJ...692..584H,2010ApJ...714..894H,2011A&A...535L...1B}
all have quiescent luminosities (well) below $10^{32}$ erg s$^{-1}$
\citep[only M15 X-3 has been detected;][]{2009ApJ...692..584H}. It
seems that such faint transients are indeed present in globular
clusters, but that their quiescent luminosities is really low. It is
worth noting that those are the only systems for which useful
constraints have been set on the quiescent properties of neutron-star
VFXTs. The Galactic disk sources are usually too much absorbed (see
also Section~\ref{forward}) for any useful constraints indicating that
globular clusters are prime targets to study quiescent
VFXTs. Sensitive monitoring programs 
\citep[see][]{wijnandsatel,altamirano2012} are needed to detect the
very-faint outbursts of VFXTs in globular clusters and combined with rapid
follow-up observations using {\it Chandra} \citep[][]{altamirano2012} to
determine the exact position, the quiescent counterpart can then be
studied in archival {\it Chandra} data or in newly proposed {\it Chandra}
observations.

It might be possible that we have underestimated the amount of energy
released per accreted nucleon \citep[see, e.g.,][]{2012PhRvC..85e5804S} and therefore the systems should be more
luminous per accreted nucleon than assumed \citep[e.g., evidence has
been reported that in some systems extra heat in shallower layers in
the crust must be liberated to explain the quiescent properties of
those systems; ][]{2009ApJ...698.1020B,2011MNRAS.418L.152D}. However,
this should also be true for the other quiescent LMXBs, which do not
require more energy per accreted nucleon to explain the base
quiescence level and most are actually too cold to be explained using
the standard model.

Therefore, we conclude that there are indeed VFXTs in globular
clusters, but that they can likely not be associated with the
brightest amongst the quiescent LMXB candidates ($>10^{33}$ erg
s$^{-1}$) and likely also not with the slightly fainter ones
($10^{32-33}$ erg s$^{-1}$).  It is more likely that those quiescent
LMXB candidates are associate with bright transients similar to the
transients IGR J17480--2446 and Swift J174805.3--244637 in Terzan 5 which were as well identified previously as quiescent LMXB candidates \citep[][]{2006ApJ...651.1098H} before they exhibited their bright outbursts
\citep[][]{2011MNRAS.412L..68D,wijnandsatel}. However, also those
transients should have low duty cycles in order to be consistent with
the observations of the lack of outbursts (either very-faint or
bright). More and more evidence becomes available that there exist a
class of transients which might have indeed very low duty cycles. This
might not be unexpected because their is a strong selection effect in
favor for discovering new transients with a high duty cycle.

\subsection{Final remarks \label{forward}}

The above discussion has demonstrated that VFXTs could be extremely
faint in quiescence but some of them could reach quiescent X-ray
luminosities of $10^{32}$ to even $10^{33}$ erg s$^{-1}$ depending on
their accretion rate history and the amount of matter accreted. So, it
would be interesting to observe quiescent VFXTs and determine their
quiescent properties. Sadly, two main factors hamper the study of the
thermal cooling emission of quiescent neutron-star VFXTs. First of
all, many quiescent LMXBs exhibit besides the thermal component a
hard, non-thermal component above 2 keV. The origin is not well
understood but its presence inhibits the most accurate study of the
thermal component and in some systems only this hard component can be
detected. In particular, there seems a trend which indicates that the
fainter a source is in quiescence, the larger the contribution is of
this non-thermal component to the 0.5--10 keV X-ray luminosity
\citep[][]{2004MNRAS.354..666J}. If this trend is also valid for the
quiescent state of VFXT, then this will make it difficult to maybe
even impossible to put significant constraints on the thermal
component in those systems.

In addition, VFXTs are difficult to discovery because their peak X-ray
luminosity is below the sensitivity limits of all-sky X-ray
instruments in orbit and therefore, most outbursts are missed. Often
only using pointed observations of more sensitive instruments, those
outbursts can be detected. But usually, those pointed observations have
a very narrow field-of-view and mostly are pointed towards the
Galactic center and the Galactic bulge. Consequently the interstellar
absorption is usually rather high for those systems with values
between $10^{22}$ cm$^{-2}$ to $10^{23}$ cm$^{-2}$. With such high
column densities and the expected low surface temperature of the
neutron stars in VFXTs, it is very difficult to impossible to detect
the thermal component. Even more so if also the non-thermal component
discussed above is present in the quiescent spectrum as well.

Despite that the situation described above looks very bleak, it might
still be possible to use quiescent VFXTs as tests for the deep-crustal
heating model. As described in Section~\ref{globulars}, many X-ray
transients are expected to be present in Galactic globular clusters
among which many VFXTs (three sources are already currently
known). Typically for many clusters the column density is not very
high and the distances are typically quite accurately known. Continued
monitoring of those clusters with sensitive X-ray instruments would be
crucial to catch VFXTs in outburst so that later they can be studied
in quiescence \citep[][]{2009ApJ...692..584H,2010ApJ...714..894H}.

When in outburst, one specific observable property can be very useful
to determine whether a VFXT is a primordial system or that it has
accreted at significantly higher rate in the past: detecting
millisecond X-ray pulsations. As discussed in Section~\ref{additional}
in the context of the accreting millisecond X-ray pulsar SAX
J1808.4--3658, detecting such millisecond X-ray pulsations basically
guarantees that the neutron star has accreted enough matter to replace
the crust. Therefore, those systems cannot be primordial (hence we
expect spin periods $>$10 ms for primordial VFXTs) and the neutron
stars systems in those systems cannot have a hybrid crust (unless in
some way the matter spins up the neutron star but is not eventually
accreted on the neutron star itself). The fact that the VFXT NGC 6440
X-2 is a millisecond X-ray pulsar with a spin frequency of 4.8 ms
\citep[][]{2010ApJ...712L..58A} indicates that this source is not a
primordial VFXT and that is faint quiescent emission
\citep[][]{2010ApJ...714..894H} cannot be due to non-standard heating
in a hybrid neutron star crust because the crust in this system should
be fully replaced and a different explanation (e.g., enhanced core
cooling) is needed.  Detailed calculations have to be performed
whether other observable properties during outbursts can discriminate
systems with a fully accreted crust with those which have a hybrid
crust (e.g., type-I X-ray bursts behavior; super-burst behavior).

Another possibility is to study those VFXTs which are active for a
very long time (years to decades, instead of weeks to months), the
so-called quasi-persistent sources, when their outbursts turn
off. Several such quasi-persistent VFXTs have been identified
\citep[see, e.g.,][albeit that those transients are close Sgr A* and
therefore have a large column density making them unsuitable for
studying the soft thermal
component]{2007A&A...468L..17D,2010A&A...524A..69D}. It has been found
that for normal transients which accrete so long that the neutron star
crust is heated considerable out of thermal equilibrium with the core
and for several months to years after the end of their outburst, the
observed quiescent temperature tracks the thermal evolution of the
crust instead of that of the core \citep[until thermal equilibrium is
reached again;
see][]{2001ApJ...560L.159W,2002ApJ...573L..45W,2003ApJ...594..952W,2004ApJ...606L..61W,2002ApJ...580..413R,2006MNRAS.372..479C,2008ApJ...687L..87C,2010ApJ...722L.137C,2010ApJ...714..270F,2011ApJ...736..162F,2011MNRAS.414L..50D,2009A&A...495..547D,2009MNRAS.396L..26D,2011MNRAS.412.1409D,2011MNRAS.418L.152D,2011A&A...528A.150D}. If a quasi-persistent VFXT can be studied right after the end of its
prolonged outburst, the observed crust cooling curve could give
significant insight into the heating and thermal processes which
occurred in the crust during the outburst. Very likely a fully
accreted crust will exhibit a different cooling profile than an hybrid
crust\footnote{We note that for the bright transient IGR J17480--2446
in Terzan 5 (which might harbor a neutron star with a hybrid crust;
see Section~\ref{additional}) the crust cooling curve has been
measured
\citep[][]{2011MNRAS.414L..50D,2011MNRAS.418L.152D,Degenaar2012} and
it did not meet the theoretical expectations. This could be due to a
neutron star with a hybrid crust \citep[see][for an in-depth
discussion]{Degenaar2012}.}. Although this assumes that still a
significant amount of heat is produced in the crust during outburst to
elevate the thermal emission above the sensitivity limits of the
current generation of X-ray instruments. 

Obtaining crust-cooling curves for neutron stars in HMXBs (i.e., after
the bright and extended type-II outbursts observed from several
neutron-star Be/X-ray transients) will also be very interesting
because it is expected that those curves could deviate significantly
from those obtained so far for the neutron-star LMXBs. Both the
effects of hybrid crusts in the neutron stars in HXMBs as well as
their high magnetic field strength might produce observational effects
on the crust cooling curves.

Primordial VFXTs will have different companions than
systems which are only in a VFXT phase but normally they are normal
transients or they used to be normal transients in the past
\citep[][]{2006MNRAS.366L..31K}. So finding the optical or IR companion star
of VFXTs would be very useful to pin-pointed the most likely accretion
rate history of those sources. Again globular clusters might be the
best places to study this because of the low absorption for many of
them compared to field targets. However, the companion star for both
the primordial and the non-primordial VFXTs can be very faint
\citep[][]{2009ApJ...692..584H,2010ApJ...714..894H} and detecting them
will remain a challenge, let alone obtaining spectral observations to
confirm the type of companion star. E-ELT using adaptive optics might
be able to detected more systems in the crowded globular clusters and
it might be able to take spectra of the brightest targets.\\


\noindent {\bf Acknowledgements.}\\ R.W. acknowledges support from a
European Research Council (ERC) starting grant. N.D. is supported by
NASA through Hubble postdoctoral fellowship grant number
HST-HF-51287.01-A from the Space Telescope Science Institute, which is
operated by the Association of Universities for Research in Astronomy,
Incorporated, under NASA contract NAS5-26555.  D.P.'s work is
supported by a grant from Conacyt (CB-2009/132400).  This research has
made use of NASA's Astrophysics Data System. We thank Craig Heinke,
Alessandro Patruno and Diego Altamirano for useful discussion. We also
thank Diego Altamirano for useful comments on a previous version of
this paper. We thank the organizers of the ``CompStar 2012: the physics
and astrophysics of compact stars'' conference on Tahiti for organizing
this nice meeting, because part of the paper was constructed there.
 
\bibliographystyle{mn2e}


\label{lastpage}
\end{document}